\documentclass[sigconf]{acmart}

\renewcommand\footnotetextcopyrightpermission[1]{} % removes footnote with conference information in first column
\setcopyright{none} % No copyright notice required for submissions
\acmConference[]
\acmYear{}

\usepackage{algorithmic, algorithm2e, color}
\usepackage{amsmath, amssymb, mathtools}
\usepackage{exscale,relsize}
\usepackage{amsfonts}
\usepackage{mdwlist}

\usepackage{graphicx}
\usepackage{caption}
\usepackage{subcaption}

\usepackage{array} 
\usepackage{url}
\usepackage{listings}
\usepackage{dsfont}
\usepackage{enumitem}
\usepackage{paralist}%%\ccsPaper{9999} % TODO: replace with your paper number once obtained
\begin{document}
\newcommand{\system}{\texttt{Thermanator}\xspace}
\newcommand{\degC}{^{\circ}C}
\newcommand{\degK}{K} % ^{\circ}K}

\newcommand{\surf}{\textsf{Shoulder-Surfing}\xspace}
\newcommand{\lunch}{\textsf{Lunch-Time}\xspace}
\newcommand{\acoust}{\textsf{Acoustic Emanations}\xspace}
\newcommand{\vibrat}{\textsf{Keyboard Vibrations}\xspace}
\newcommand{\attack}{\textsf{Thermanator}\xspace}

\newcommand{\surfa}{\textsf{Shoulder-Surfing Attacks}\xspace}
\newcommand{\luncha}{\textsf{Lunch-Time Attacks}\xspace}
\newcommand{\attacka}{\textsf{Thermanator Attacks}\xspace}
\newcommand{\acousta}{\textsf{Acoustic Emanations Attacks}\xspace}
\newcommand{\vibrata}{\textsf{Vibration Attacks}\xspace}
\title{\system: Thermal Residue-Based Post Factum Attacks On Keyboard Password Entry}

	\author{Tyler Kaczmarek}
	\affiliation{ UC Irvine}
	\email{tkaczmar@uci.edu} 

	\author{ Ercan Ozturk}
	\affiliation{ UC Irvine}
	\email{ercano@uci.edu}

	\author{ Gene Tsudik}
	\affiliation{ UC Irvine}
	\email{gene.tsudik@uci.edu}
\begin{abstract}
As a warm-blooded mammalian species, we humans routinely leave thermal residues on various objects with which 
we come in contact. This includes common input devices, such as keyboards, that are used for 
entering (among other things) secret information, such as passwords and PINs. Although thermal 
residue dissipates over time, there is always a certain time window during which thermal energy 
readings can be harvested from input devices to recover recently entered, and potentially sensitive,
information. 

To-date, there has been no systematic investigation of thermal profiles of keyboards, 
and thus no efforts have been made to secure them. This serves as our main motivation 
for constructing a means for password harvesting from keyboard thermal emanations. 
Specifically, we introduce \system, a new post factum insider attack 
based on heat transfer caused by a user typing a password on a typical external keyboard. 
We conduct and describe a user study that collected thermal residues from $30$ 
users entering $10$ unique passwords (both weak and strong) on $4$ popular commodity keyboards. 
Results show that entire sets of key-presses can be recovered by non-expert users as late as \textbf{$30$ seconds} 
after initial password entry, while partial sets can be recovered as late as \textbf{$1$ minute} after entry. Furthermore, 
we find that Hunt-and-Peck typists are particularly vulnerable.  We also discuss some \attack mitigation strategies.

%%GTS: not sure about this... Perhaps too much selling?
%%TK: (3) is a bit much, and we never touch on abandoning passwords anywhere else (1) and (2) seem reasonable (especially (2))
The main take-away of this work is three-fold: (1) using external keyboards 
to enter (already much-maligned) passwords is even less secure than previously recognized, (2) post factum (planned or impromptu) 
thermal imaging attacks are realistic, and finally (3) perhaps it is time to either stop using keyboards for password entry,
or abandon passwords altogether.
\end{abstract}

\maketitle

%\keywords{Side-Channel; Thermal Images; Lunchtime Attack }

\section{Introduction}
\label{sec:intro}
Insider attacks are very common, estimated to account for ${\approx}28\%$ of all 
electronic crimes in industry \cite{mickelberg2014us}. This includes some high-profile attacks, 
such as the 2014 Sony hack \cite{robb2014sony}. At the same time, it is well known that security 
of a system is based on its weakest link. Furthermore,  it is often assumed that involvement of a fallible 
(or simply gullible) human user corresponds to this weakest link, e.g., as in \surf and \lunch attacks. 
However, other insider attacks that focus on stealing passwords by compromising the user environment, 
e.g., \acoust \cite{asonov2004keyboard,zhuang2009keyboard,compagno2017don}
or \vibrat \cite{owusu2012accessory}, show that the weakest link is a consequence of a law of Physics.    
However, such insider attacks must occur instantaneously, in real time, in order to succeed. In other words, to
exploit them, the adversary must be able to record the environment as the user is entering a password. 
Real-time adversarial presence (whether in person or via a nearby compromised recording device)
raises the bar for the attack. This prompts the question: \\
{\em Are there any observable physical effects of password entry that linger and can therefore be 
collected {\bf afterwards}?}

\subsection{Heat Transfer \& Thermal Emanations}
Any time two objects with unequal temperatures come in contact with each other, an 
exchange of heat occurs. This is unavoidable. Being warm-blooded, human beings naturally prefer
environments that are colder than their internal temperature. Because of this heat disparity, 
it is inevitable that we leave thermal residue on numerous objects that we routinely touch, especially, with 
with bare fingers. 
Furthermore, it takes time for these heated objects to cool off and lose heat energy imparted by 
human contact. It is both not surprising and worrisome that this includes our interactions with 
keyboards that are used for entering sensitive private information, such as passwords. 

Based on this observation, we consider a mostly unexplored attack space where heat 
transfer and subsequent thermal residue can be exploited by a clever adversary to steal 
passwords from a keyboard some time after it was used for password entry. The main
distinctive benefit of this attack type is that adversary's real time presence is not required.
Instead, a successful attack can occur with after-the-fact adversarial presence:  
as our results show, many seconds later.

While there has been some prior work on using thermal emanations to crack PINs, mobile 
phone screen-locks and  opening combinations of vaults/safes \cite{SafeCracking,andriotis2013pilot, 
abdelrahman2017stay,mowery2011heat}, this work represents the first comprehensive investigation of 
human-based thermal residues and emanations of external computer keyboards.

\subsection{Expected Contributions}
In this paper, we propose and evaluate a particular human-based side-channel attack class, called \attack. This attack class 
is based on exploiting thermal residues left behind by a user (victim) who enters a password using
a typical external keyboard. Shortly after password entry, the victim either steps away inadvertently, or is drawn away
(perhaps as a result of being prompted by the adversary) from the personal workplace. Then, the adversary captures 
thermal images of the victim keyboard. We examine the efficacy of \attacka for a moderately sophisticated adversary 
equipped with a mid-range thermal imaging camera.  The goal of the attack is to learn information about the victim password.

To confirm viability of \attacka, we conducted a rigorous two-stage user study. The first stage collected password entry 
data from 31 subjects using 4 common keyboards. In the second stage, 8 non-expert subjects acted as adversaries and 
attempted to derive the set of pressed keys from the thermal imaging data collected in the first stage. Our results show that 
even novice adversaries can use thermal residues to reliably determine the entire set of key-presses {\bf up to 30 seconds} 
after password entry. Furthermore, they can determine a partial set of key-presses as long as a full minute after password 
entry. We provide a thorough discussion of the implications of this study, and mitigation techniques against \attacka. 

Furthermore, in the course of exploring \attacka, we introduce a new post factum adversarial model. We comprehensively compare 
this model with those of other insider attacks that target user behavior and physical properties, such as \lunch, \surf, and \acoust  
attacks. In doing so, we focus on attack characteristics, such as: goals, timeline and equipment required by the adversary. 

\subsection{Organization}
Section \ref{sec:bg} provides background on thermodynamic concepts,  modern keyboards and thermal cameras. 
Section \ref{sec:advers} describes assumed \attacka and 
adversarial models. Section \ref{sec:method} describes our methodology, apparatus and subject recruitment. Study results are 
presented in Section \ref{sec:res} and their implications are discussed in Section \ref{sec:dis}. We then compare and contrast 
\attack with other insider attacks in Section \ref{sec:compare}. Section \ref{sec:rw} discusses related work. 
We conclude the paper with directions for future work in Section \ref{sec:conc}.

\section{Background}
\label{sec:bg}
In this section we provide some background material on physical interactions that describe thermal phenomena observed 
in our experiments. We start with a glossary of terms, then describe the form factor and material composition of 
modern 104-key "Windows" keyboards and finish with certain Physics concepts used in the rest of the paper. 
Given familiarity with elements of Conductive Heat Transfer and Newton's Law of Cooling, 
Sections \ref{subsec:glossary}, \ref{subsec:conduct}, and \ref{subsec:convect} can be skipped with no loss of continuity.

\subsection{Basic Thermal Terminology}
\label{subsec:glossary}
\begin{compactitem}
\item Joule (J) - Unit of energy Corresponding to $1$ Newton-Meter ($N\over{m}$)
\item Kelvin ($\degK$) -- Base unit of temperature in Physics. 
The temperature T in Kelvin (\degK) minus $273.15$ yields the corresponding temperature in degrees Celsius ($\degC$).
\item Watt (W) -- Unit of work corresponding to 1 Joule-Second: ($J\over{s}$)
\item Conduction --  Transfer of Thermal Energy caused by two objects in physical contact that are at different Temperatures.
\item Convection --  Transfer of Thermal Energy caused by submerging an object in a fluid.
\item Heat Transfer Coefficient - Property of a fluid that determines rate of convective heat flow. 
Expressed in Watts per square meter Kelvin: $W\over{m^2} {\degK}$
\item Specific Heat -- Amount of Thermal Energy in Joules that it takes to increase temperature of $1$kg 
of material by $1\degK$.  Expressed in Joules over kilograms degrees Kelvin: $J\over{kg}\degK$.
\item Thermal Conductivity --  Rate at which Thermal Energy passes through a material. Expressed in 
Watts per meters Kelvin: $W\over{m}\degK$
%TK the "m" here is meters, not mili kelvin, hence why all refs. to mili kelvin have been removed
\item Thermal Energy -- Latent energy stored in an object due to heat flowing into it.
\item Thermal Source -- Object or material that can internally generate Thermal Energy such that it can stay at constant 
temperature during a thermal interaction, e.g., a heat pump.
\end{compactitem}

\subsection{Heating via Thermal Conduction}
\label{subsec:conduct}
Thermal Conduction is transfer of heat between any two touching objects of different temperatures. It is expressed as the 
movement of heat energy from the warmer to the cooler object. We are concerned with transfer of energy from a human 
fingertip to a pressed keycap. This transfer is governed by Fourier's Law of heat conduction which states that: 
\begin{quote}\em
Heat transfer between two objects can be modeled by the equation: 
$q = {\mathcal{K}A(T_1-T_2)t \over d}$, where $\mathcal{K}$ is thermal conductivity\footnote{$\mathcal{K}$ should not be
confused with $\degK$ -- degrees Kelvin.}
of the object being heated, $A$ is area of contact, $T_1$ is  initial temperature of the hotter object, $T_2$ is initial temperature of 
the cooler object,  $t$ is time, and $d$ is the thickness of the object being heated.  
\end{quote}
The relationship between an object's heat energy and its temperature is governed by the object's mass and 
specific heat, as dictated by the formula: $q = c m \Delta T$, where $q$ is total heat energy, $c$ is object's specific heat, 
$m$ is object's mass and $\Delta T$ is change in temperature.

We consider the human body to be a thermal source, and we assume that any change in the fingertip temperature 
during the (very short) fingertip-keycap contact period is negligible, due to internal heat 
regulation \cite{dai2004comparison}. Furthermore, we assume that:
\begin{compactitem}
\item Average area of an adult human fingertip is $400 mm^{2}$  \cite{peters2009diminutive}.
\item Average human skin temperature is $307.15\degK$ ($= 34\degC$) \cite{burton1939range}. 
\item Average duration of a key-press is $0.28$s \cite{sauro2009estimating}. 
\item Keyboard temperature is the same of that as that of the air, which, for a typical office, 
is OSHA\footnote{OSHA = Occupational Safety and 
Hazards Administration, a United States federal agency.}-recommended 
$294.15\degK$ ($=21\degC$) \cite{occupational1999osha}. 
\end{compactitem}
Therefore, for variables mentioned above, we have: 

\begin{center}
\fbox{
$$ \bf\footnotesize
$\mathcal{K}$=0.25, \; A=0.00024025, \; T1=34, \; T2=21, \;  t=0.28, \; \mbox{and} \; d=0.0015
$$
}
\end{center}

Plugging these values into Fourier's Law, we get: 
$$
q = {(0.25)(0.00024025)(34-21)(.28) \over 0.0015}
$$
which yields 
total energy transfer: $q=0.1458$J.  We then use total energy $q$ in the specific heat equation to determine total temperature 
change: $0.1458=(1000)(0.0004716)\Delta T$. This gives us a total temperature change of $\Delta T = 0.3092$. 
Therefore, we conclude that the average human fingertip touching a keycap 
at the average room temperature results in the keycap heating up by $0.3092\degK$.

\subsection{Cooling via Thermal Convection}
\label{subsec:convect}
After a keycap heats up as a result of conduction caused by a press by a warm(er) human finger, it begins to cool off 
due to convective heat transfer with the air in the room. Convection is defined as the transfer of heat resulting from the 
internal current of a fluid, which moves hot (and less dense) particles upward, and cold (and denser) particles -- downward. 
This interaction is governed by Newton's Law of Cooling. Its particulars are impacted by the shape and position of the 
heated object. In our case, there is a plane surface\footnote{The actual keycap surface can be slightly concave.} 
facing towards the cooling fluid (i.e., a keycap directly exposed to ambient air)
which is described by the formula: 
$$
T(t) = T_{s} + (T_{0} -T_{s})e^{-\kappa t}
$$ 
where $T(t)$ is temperature at time $t$, $T_s$ is temperature of ambient air, 
$T_0$ is initial object temperature, and $\kappa$ is the cooling constant of still (non-turbulent) air 
over a $0.00024025m^2$ plane.

This comes with the additional intuitive notion that a surface convectively cools quicker when the 
temperature difference between the heated object and the fluid is higher. Similarly, it cools slower when 
the temperature difference is smaller. Finally, Newton's Law of Cooling is asymptotic, and cannot be used 
to find the time at which the object reaches the exact temperature of the ambient fluid. 
Thus, instead of finding the time when the temperatures are equal, we determine the time when the temperature 
difference falls below an acceptable threshold, which we set at $0.04\degK$. Plugging this into Newton's 
Law of Cooling results in: 
$$ 
t = - {ln({0.3092\over 0.04}) \over 0.037}
$$
which yields $t= 55.7$ for total time for a pressed key to cool down to the point where it is indistinguishable from room temperature.

\subsection{Modern Keyboards}
\label{subsec:composition}
\begin{figure}[ht!]
\fbox{\centering
\includegraphics[height=1.5in,width=0.95\columnwidth]{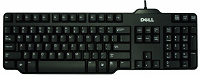} }
\caption{\small{Typical "Windows"-style Keyboard.}}
\label{fig:board}
\end{figure}

Most commodity external keyboard models are of the 104-key "Windows" variety, shown in Figure \ref{fig:board}. 
On such keyboards, the distance between centers of adjacent keys is
about $19.05$mm, and a typical keycap shape is an  $\approx[15.5mm~x~15.5mm~x~1.5mm]$ rectangular prism, with 
an average travel distance of $3.55mm$ \cite{noyes1983qwerty}; see Figure \ref{fig:keycap}. 
All such keyboards are constructed out of Polybutylene Terephthalate (PBT) with density of $1.31g/cm^{3}$ , 
resulting in an average keycap mass of $.4716g$ \cite{pyda2004heat}.  PBT generally has the following characteristics: 
specific heat = $1,000{J \over kg \degK}$ and thermal conductivity = $0.274{W\over m\degK}$ \cite{pyda2004heat}.
\begin{figure}[ht!]
\fbox{\centering
\includegraphics[height=1.3in,width=0.7\columnwidth]{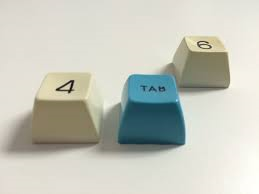} }
\caption{\small{Typical Keycap Profile.}}
\label{fig:keycap}
\end{figure}

\subsection{Thermal Cameras}
\label{subsec:FLIRS}
In the past few years, many niche computational and sensing devices have moved from Hollywood-style fantasy
into reality. This includes thermal imagers or cameras. In order to clarify their availability to individuals (or agencies) 
at different levels of sophistication, we provide the following brief comparison of several types of readily-available 
FLIR: {\bf F}orward-{\bf L}ooking {\bf I}nfra-{\bf R}ed devices. (See: \url{https://www.flir.com/products} for full product specifications.)
In the rest of the paper, we use the following terms interchangeably: FLIR device, thermal imager and thermal camera.
\begin{figure}[H]
\fbox{\centering
\includegraphics[height=2.5in,width=\columnwidth]{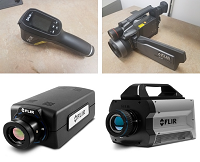} }
\caption{\small{FLIR Devices / Thermal Imagers: TG165(top left,) SC620(top right,) A6700sc (bottom left,) and X8500sc (bottom right).}}
\label{fig:keycap}
\end{figure}
\begin{compactenum}
\item[TG165] -- Price: About US\$$300$. Thermal Sensitivity: $0.15$K. Thermal Accuracy:  
$\pm 1.5$K or $1.5\%$ of reading. Resolution:$50x80$. Image Capture: Manual, $1$ image at a time. Video Capture: None
\item[SC620] -- Price: About US\$$1500$ (used). Thermal Sensitivity: $0.04$K Thermal Accuracy: 
$\pm 2$K or $2\%$ of reading. Resolution: $640x480$. Image Capture: Automatic, programmable to capture 
images by timer, or when specific criteria are met, at maximum rate of $1$ image per second. Video Capture: None.
\item[A6700sc] -- Price: About US\$$25,000$. Thermal Sensitivity: $0.018$K Thermal Accuracy: 
$\pm 2$K or $2\%$ of reading. Resolution: $640x512$. Image Capture: Automatic, programmable to capture  
images by timer or when specific criteria are met, at up to $100$fps. Video Capture: High speed, up to 
$100$fps.
\item[X8500sc] -- Price: About US\$$100,000$. Thermal Sensitivity: $0.02$K: Thermal Accuracy: 
$\pm 2$K  or $2\%$ of reading. Resolution: $1280x1024$ Image Capture: Automatic, programmable to capture 
images by timer or when specific criteria are met, at up to $180$fps. Video Capture: High speed, up to 
$180$fps.
\end{compactenum}
Obviously, a sufficiently motivated organization or a nation-state could easily obtain thermal 
imagers of the highest quality (and price), we assume that the anticipated adversary is of a mid-range 
sophistication level, i.e., capable of acquiring a device of the type exemplified by SC620.

\section{Adversarial Model \& Attacks}
\label{sec:advers}
This section describes the adversarial model for \attacka.
\begin{figure*}[ht!]
	\centering
	\begin{subfigure}{0.3\textwidth}
		\includegraphics[width=\textwidth]{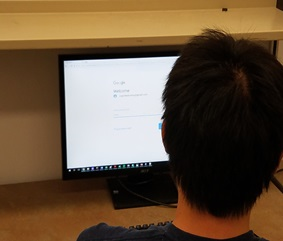}
		\caption{STEP 1: Victim Enters Password} 
	\end{subfigure}
	\begin{subfigure}{0.3\textwidth} 
		\includegraphics[width=\textwidth]{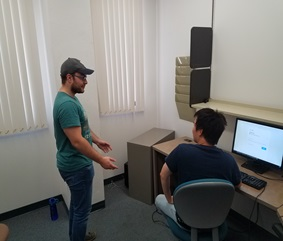}
		\caption{STEP 2: Victim Leaves} 
	\end{subfigure}
	\begin{subfigure}{0.3\textwidth} 
		\includegraphics[width=\textwidth]{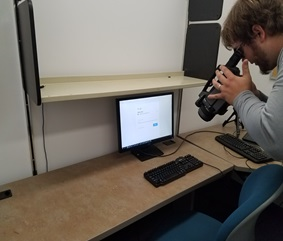}
		\caption{STEP 3: Thermal Residues Captured} 
	\end{subfigure}
	\caption{An Example \attack Attack} 
	\label{attack}
\end{figure*}

\subsection{Physical Premise}
As mentioned in Section \ref{sec:bg}, Fourier's Law states that contact between any two objects with unequal temperatures 
results in transfer of heat energy from the hotter to the cooler object. It is reasonable to assume that the typical office 
environment has the ambient temperature within the OSHA-recommended range of $293.15-298.15\degK$ 
(=$20-25\degC$) \cite{occupational1999osha}. 
In that setting, the average human hand is expected to conductively transfer an observable amount of heat to the 
ambient-temperature keyboard. Consequently, a bare-fingered human typist can not avoid leaving thermal residue on a keyboard. 
This physical interaction can be abused by the adversary in order to harvest the thermal residue of a victim who recently 
used a keyboard to enter potentially sensitive information, e.g., a password. This forms the premise for \attacka.

\subsection{\attack}
\attack is a distinct type of insider attack, where a typical attack scenario proceeds as follows:
\begin{description}
\item[STEP 1:] The victim uses a keyboard to enter a genuine password, as part of the log-in (or session unlock) procedure.
\item[STEP 2:] Shortly thereafter, the victim either: (1) willingly steps away, or (2) gets drawn away, from the workplace.
\item[STEP 3:] Using thermal imaging (e.g., photos taken by a commodity FLIR camera) the adversary harvests thermal 
residues from the keyboard.
\item[STEP 4:] At a later time, the adversary uses the ``heat map'' of the images to determine recently pressed keys. 
This can be done manually (i.e., via visual inspection) or automatically (i.e., via specialized software). 
\item [\underline{REPEAT:}] The adversary can choose to repeat STEPS [1-4] over multiple sessions. 
\end{description}
The two options in \textbf{STEP 2} correspond to two attack sub-types: \emph{opportunistic} and \emph{orchestrated}.
In the former, the adversary patiently waits for the situation described in \textbf{STEP 2} case (1) to occur. Once the victim leaves 
(on their own volition) shortly after password entry, the adversary swoops in and collects thermal residues. This strategy
is similar to \luncha. In an \emph{orchestrated} attack,
instead of waiting for the victim to leave, the adversary uses an accomplice to draw the victim away shortly
after password entry, as in \textbf{STEP 2} case (2).

\section{Methodology}
\label{sec:method}
In this section we describe of the experimental apparatus, procedures, and subject recruitment methods.

\subsection{Apparatus}
\label{subsec:app}
\begin{figure*}[ht!]
	\centering
	\begin{subfigure}{0.3\textwidth}
		\includegraphics[width=\textwidth]{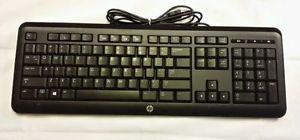}
		\caption{HP SK-2023} 
	\end{subfigure}
	\begin{subfigure}{0.3\textwidth} 
		\includegraphics[width=\textwidth]{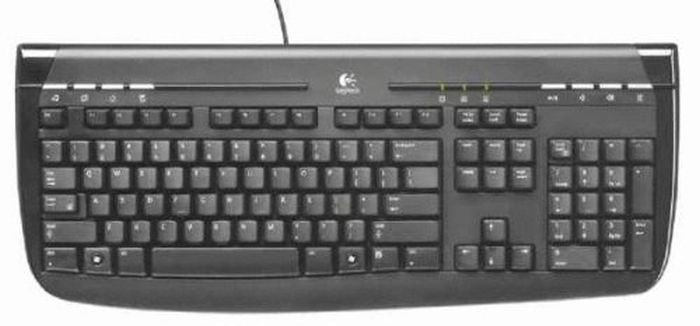}
		\caption{Logitech Y-UM76A.} 
	\end{subfigure}
	\begin{subfigure}{0.3\textwidth} 
		\includegraphics[width=\textwidth]{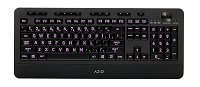}
		\caption{AZiO Prism KB507 (backlit).} 
	\end{subfigure}
	\caption{Keyboards} 
	\label{kboards}
\end{figure*}

\begin{figure}[th!]
\fbox{\centering
\includegraphics[height=2.0in,width=\columnwidth]{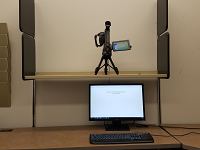} }
\caption{\small{SC620 Apparatus Setup}}
\label{fig:620Setup}
\end{figure}

The experimental setup was designed to simulate a typical office setting. It was located in a dedicated 
office in a research building of a large university. Since experiments were conducted during the academic year,
there was always some (though not excessive) amount of typical busy office-like ambient noise. Figure \ref{fig:620Setup} 
shows the setup from the subject's perspective.
Equipment used in the experiments consisted of the following readily available (off-the-shelf) components:
\begin{compactenum}
\item[I.] FLIR Systems SC620 Thermal Imaging Camera\footnote{see: \url{http://www.FLIR.com} for a full specification.} 
This camera was perched on a tripod $24"$\ above the keyboard.
\item[II.] Four popular and inexpensive commodity computer keyboards:
(a) Dell SK-8115,
(b) HP SK-2023
(c) Logitech Y-UM76A, and
(d) AZiO Prism KB507.
The first (Dell) is shown in \ref{fig:board} above, and the other three -- in Figure \ref{kboards}.
\end{compactenum}

The thermal camera was chosen to be realistic for a moderately sophisticated and determined adversary. We assume
this type of adversary to be an individual, i.e., not an intelligence agency or a powerful criminal organization. 
FLIR SC620 Thermal Imager costs approximately $\mbox{US}\$1,500$ used. (This model is about 6-7 years old.) 
It automatically records images at the resolution of $640x480$ pixels, with $1Hz$ frequency. Its thermal sensitivity is $0.04\degK$. 

The four keyboards were chosen to cover the typical range of manufacturers represented in an average workplace. 
Dell, HP and Logitech keyboards are popular default keyboards included in new computer orders from major 
PC, desktop, and workstation manufacturers. Each costs $\approx\mbox{US}\$20$. 
Meanwhile, Azio Prism is a popular low-cost and independently manufactured keyboard that can be 
easily obtained on-line e.g., from Amazon; it costs $\approx\mbox{US}\$25$.

\begin{figure}[th!]
\fbox{\centering
\includegraphics[width=\columnwidth]{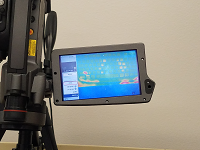} }
\caption{\small{Example of Thermal Emanations being Recorded.}}
\label{fig:620Data}
\end{figure}

\subsection{Procedures}
\label{subsec:proc}
\system was evaluated using a two-stage user study. The first stage was conducted to collect thermal 
emanation data, and the second -- to evaluate efficacy of \attacka. A given subject only participated in a 
single stage.

\subsubsection{Stage One: Password Entry}
\begin{figure}[ht!]
\fbox{\centering
\includegraphics[height=2.8in,width=\columnwidth]{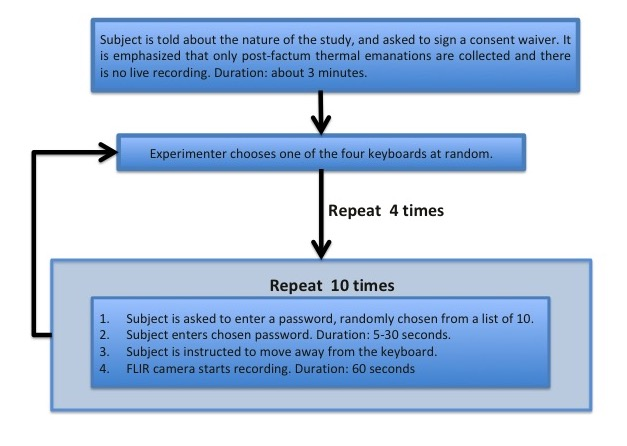} }
\caption{\small{Experiment Stage One: Flowchart}}
\label{fig:flowchart}
\end{figure}

Recall that \attack 's goal is to capture thermal residues of subjects {\bf after} keyboard password entry.
This is accomplished by FLIR SC620 taking a sequence of images (60 total), once per second for a total
of one minute after initial password entry. The first stage is shown in Figure \ref{fig:flowchart}.

Each subject entered $10$ passwords on $4$ keyboards and each entry was followed by one minute of keyboard recording 
($60$ successive images) by the FLIR. Each subject entered a total of $40$ passwords and every entry took, on average,
between $10$ and $20$ seconds. The total duration of the experiment for a Stage 1 subject ranged between $50$ and $60$ minutes, 
based on the individual's typing speed and style. 
Both keyboards and passwords were presented to each subject in random order, in an attempt to negate
any side-effects due to subject training or familiarity with the task. 

We selected 10 passwords that included both "insecure" and "secure" categories. The former passwords were culled from 
the top 100 passwords by popularity that adhere to common password requirements, such as Gmail 
\footnote{see: \url{https://support.google.com} for details}. Whereas, "secure" passwords were created by randomly 
generating 8-, 10-, and 12-character strings of lower/uppercase letters as well as numbers and symbols 
that adhere to Gmail restrictions. Our selection criteria resulted in the following 10 candidate passwords: 
\begin{compactitem}
\item {\bf [Insecure]:} "password", "12345678", "football", "iloveyou", "12341234", "passw0rd", and "jordan23",
\item {\bf [Secure]:} "jxM\#1CT[", "3xZFkMMv|Y", and \\"6pl;0>6t(OvF".
\end{compactitem}

\subsubsection{Stage Two: Data Inspection}
\begin{figure}[h]
\fbox{\centering
\includegraphics[height=1.8in,width=\columnwidth]{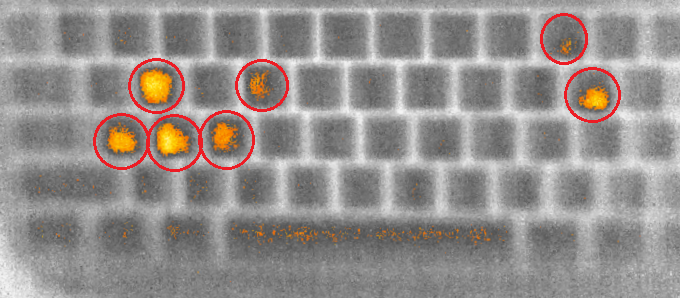} }
\caption{\small{Thermal image of "passw0rd" $20$seconds after entry.}}
\label{fig:pw20}
\end{figure}
The second stage of the experiment has subjects act as adversaries conducting \attacka. 
Subjects were shown images obtained from the first stage of the experiment, e.g., Figure \ref{fig:pw20}, 
and were instructed to identify the "lit" regions. Each subject was shown $150$ recordings of password entries 
in random order. On average, a subject could process a single recording in $45-60$ seconds. Total time for each 
Stage 2 subject varied in the range of $100-130$ minutes. 

\subsection{Subject Recruitment Procedure}
Subjects were recruited from the (student body of a large public University using a unified Human Subjects Pool 
designated for undergraduate volunteers seeking to participate in studies such as ours. Subjects were compensated 
with course credit. Because of this, overwhelming majority of subjects were of college age: $18--25$. The subject gender
breakdown was: $16$ male and $15$ female.  

All experiments were authorized by the Institutional Review Board (IRB) of the authors' employer, 
well ahead of the commencement of the study. The level of review was: Exempt, Category II. 
No sensitive data was collected during the experiments and minimal  identifying information was retained. 
In particular, no subject names, phone numbers or other personally identifying  information (PII) was collected.
All data is stored pseudonymously.

\section{Results}
\label{sec:res}
\begin{figure}[h]
\fbox{\centering
\includegraphics[height = 1.5in, width=.9\columnwidth]{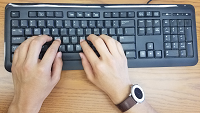} }
\caption{\small {A Touch Typist's Hands Perched on the Home-Row }}
\label{fig:hr}
\end{figure}

We now describe the results of Stage 2 analysis of thermal images obtained in Stage 1. 
We divide it into two categories: 
\begin{compactitem}
\item  Hunt-and-Peck Typists --- `those who {\bf do not} rest their fingertips on, or hover their fingers just over, 
the home-row of keys: 
\item Touch Typists -- those whose fingertips routinely hover over, or lightly touch, the home-row, 
as shown by Figure \ref{fig:hr}. 
\end{compactitem}
As it turns out, our study results indicate that the category of the typist is the most influential 
factor for the quality thermal imaging data. For each category, 
we separately analyze "secure" and "insecure" passwords types

\begin{figure}[h]
\fbox{\centering
\includegraphics[height = 2.0in, width=\columnwidth]{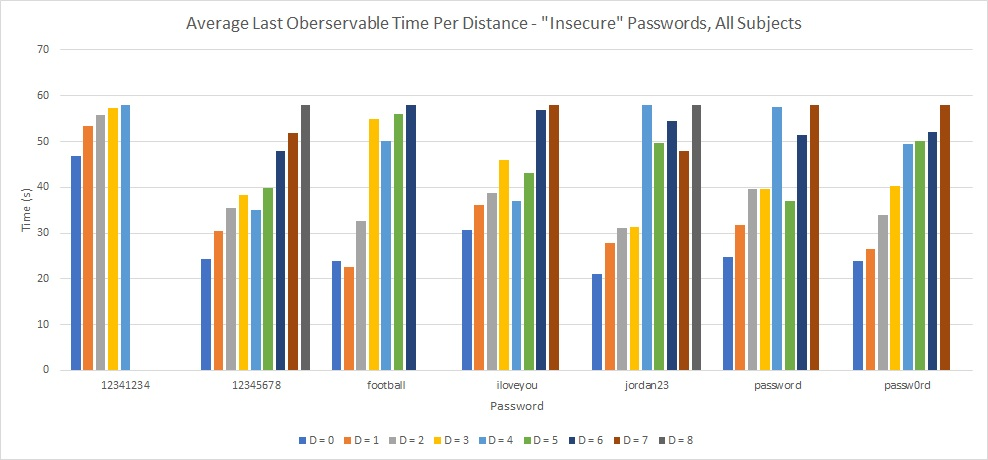} }
\caption{\small {ID Rates for All Subjects: "Insecure" Passwords }}
\label{fig:allinsec}
\end{figure}

\begin{figure}[h]
\fbox{\centering
\includegraphics[height = 2.0in, width=\columnwidth]{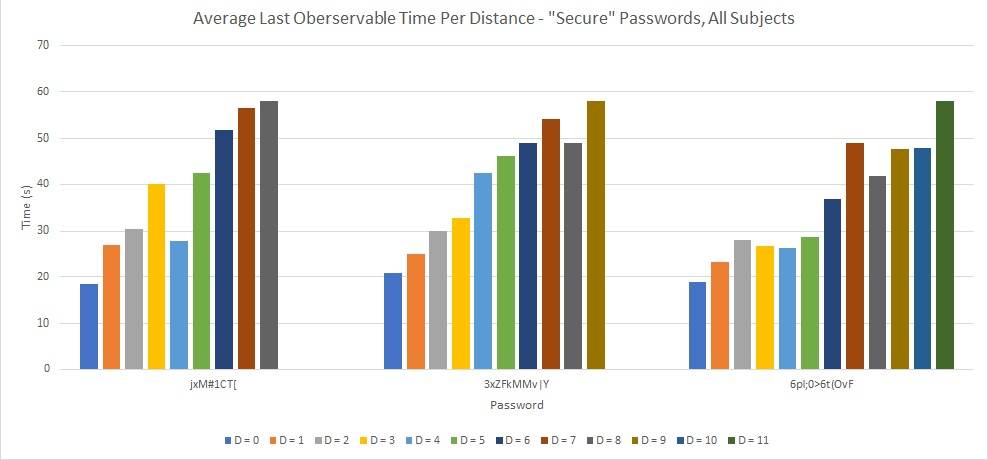} }
\caption{\small {ID Rates for All Subjects: "Secure" Passwords}}
\label{fig:allsec}
\end{figure}

For full context, aggregate results (identification rates) from the entire subject population are shown in Figures \ref{fig:allinsec} and 
\ref{fig:allsec}; they correspond to stage 2 subjects' analysis of "insecure" and "secure" passwords, respectively. 
In each graph, "d = 0" refers to average latest time when stage 2 subjects could correctly identify every keystroke 
of the entered password, while "d = 1" denotes average latest time when subjects could identify all-but-one keystroke, 
"d = 2" denotes the average latest time when subjects could identify all-but-two keystrokes and so on.

\subsection{Hunt-and-Peck Typists}
\label{subsec:handp}
Our analysis of Hunt-and-Peck typists was straightforward. Because these typists do not rest their fingertips 
on (or hover right above) the keyboard home-row, it is readily apparent that each bright spot on the thermal image 
corresponds to a key-press. However, as discussed below, we encountered some challenges with "secure" passwords.

\subsubsection{Insecure Passwords}
\begin{figure}[h]
\fbox{\centering
\includegraphics[height = 2.0in, width=\columnwidth]{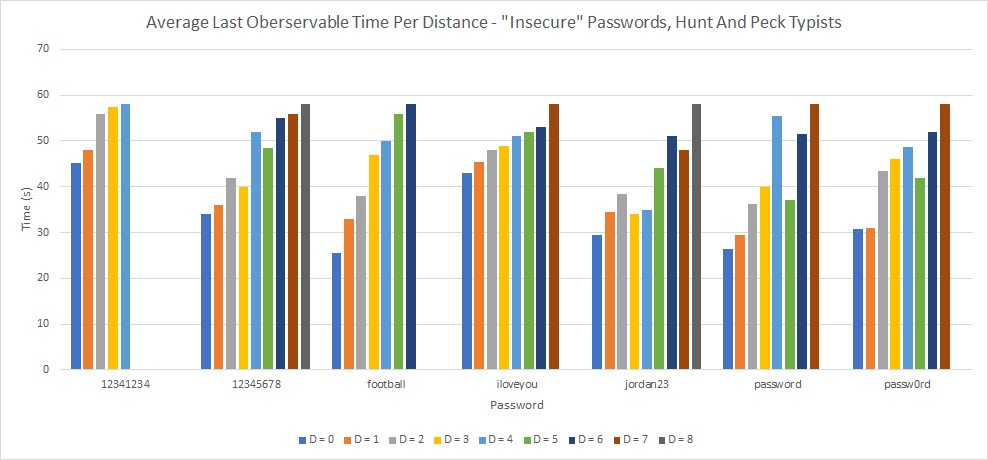} }
\caption{\small {ID Rates for Hunt-and-Peck: "Insecure" Passwords }}
\label{fig:handpinsec}
\end{figure}
As Figure \ref{fig:handpinsec} shows, analysis of Hunt-and-Peck typists entering "insecure" passwords is straightforward. 
In fact, in the best-case of "12341234" subjects could correctly recall every keystroke, on average, $45.25$ seconds after 
entry. Even the weakest result, "football" was fully recoverable $25.5$ seconds later, on average. This is in line with 
conventional thought. Hunt-and-Peck typists typically only use their forefingers to type. Because of this, they make contact 
with a larger finger over a large surface area. Also, since Hunt-and-Peck typists are generally less skilled, they take 
longer for each keystroke, resulting in longer contact time. These two factors combined yield high-quality thermal residue 
for \attacka . 

\subsubsection{Secure Passwords}
\begin{figure}[h]
\fbox{\centering
\includegraphics[height = 2.0in, width=\columnwidth]{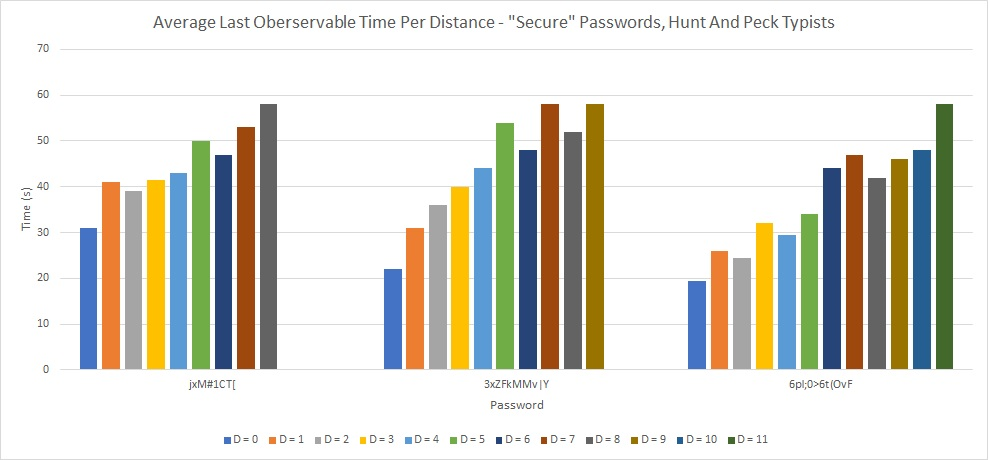} }
\caption{\small {ID Rates for Hunt-and-Peck: "Secure" Passwords}}
\label{fig:handpsec}
\end{figure}
"Secure" passwords are more challenging to analyze. As shown in Figure \ref{fig:handpsec} full recall was possible, on average, 
up to $31$ seconds after recording started, in the best case, and $19.5$ seconds, in the worst case. Performance of stage 2 
subjects was uniform in terms of password length: the shortest password was the easiest to analyze correctly. 
Anecdotally, this is not surprising. It was quite common for Hunt-and-Peck typists to look back and forth between the characters
of a relatively complex "secure" passwords, and their keyboards. This resulted in longer completion times, 
which left longer time for keycaps to cool off before recording began.

\subsection{Touch Typists}
\label{subsec:touch}
Analyzing data from Touch typists was a challenge for stage 2 subjects. Since a typical Touch typist's fingers are 
constantly in contact with (or in very close proximity of) the home-row of the keyboard, there are two incidental 
sources of thermal noise. First, there is thermal residue on the 2 groups of 4 home-row keys: "asdf" and "jkl;" 
which results from the typist's fingertips. However, whenever typist's ingers rest on the keyboard for a long time,
additional observed effects occur outside (though near) the home-row, on the following keys:

\fbox{{\centering\noindent\bf
{\tt "qwertgvcxz"} on the left, {\tt "][poiuhnm,./"} on the right
}}
\\
Even though this secondary thermal residue was not as drastic as that on the home-row, it had a more pronounced 
effect on stage 2 subjects. In many cases, a subject was uncertain whether a key was lit on the 
thermal image because it was actually pressed, or because it was simply close to the home-row. 
This uncertainty in turn led to mis-classification of some keys as unpressed. Also, mis-classification of 
home-row keys as pressed keys was not counted in the distance. We justify this choice in Section \ref{sec:dis}.

\subsubsection{Insecure Passwords}
\begin{figure}[h]
\fbox{\centering
\includegraphics[height = 2.0in, width=\columnwidth]{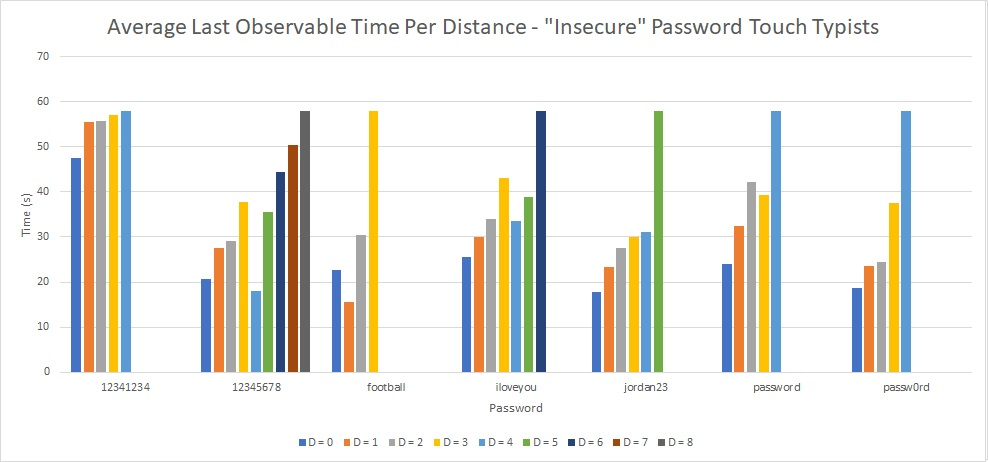} }
\caption{\small{ID Rates for Touch Typists: "Insecure" Passwords }}
\label{fig:touchinsec}
\end{figure}
While more difficult than analysis of "insecure" password for Hunt-and-Peck typists, phase 2 subjects has 
moderate success analyzing Touch typists entering "insecure" passwords. As Figure \ref{fig:touchinsec} 
shows, the best average time for full recall was for password: "12341234" at $47.6$ seconds, and the worst 
was for "jordan23", at $17.8$\ seconds. This follows the notion that stage 2 subjects were hesitant to classify 
home-row-adjacent key-presses, e.g., "o", "r" and "n" in "jordan23". Furthermore, this supports the notion that a 
simple, repeated password such as "12341234" leaves ideal thermal residue. Since each key is repeated, 
it is analogous to each key being pressed once for twice as long. This results in twice as much thermal energy 
being transferred from the fingertip to the keycap.

\subsubsection{Secure Passwords}
\begin{figure}[h]
\fbox{\centering
\includegraphics[height = 2.0in, width=\columnwidth]{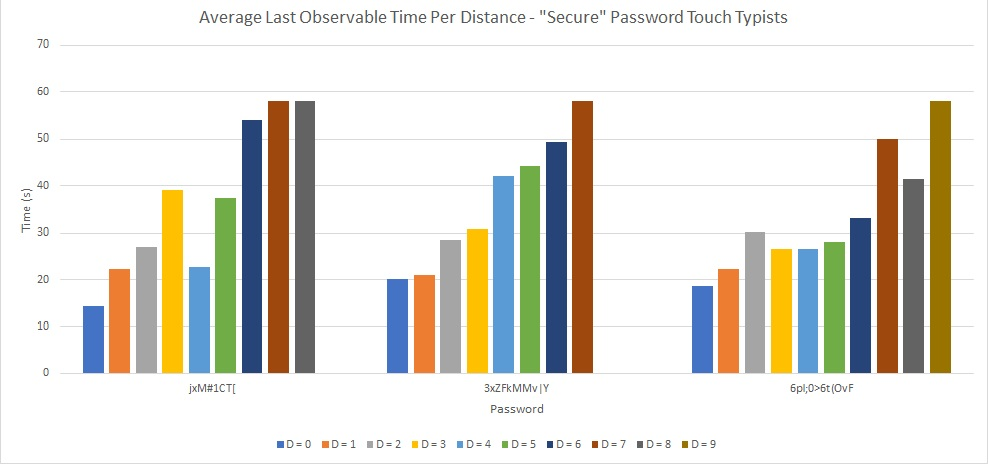} }
\caption{\small{ID Rates for Touch Typists: "Secure" Passwords}}
\label{fig:touchsec}
\end{figure}
Touch typists entering "secure" passwords were the most difficult for the stage 2 subjects to analyze. 
As shown in Figure \ref{fig:touchsec}, full recall was only possible, on average, within the first $14.33--18.5$ 
seconds. Surprisingly, the password with the smallest window for full recall was "jxM\#1CT[".  We believe 
that many phase 2 subjects were hesitant to classify home-row-adjacent keys in this password as keystrokes
(as opposed to thermal noise). This might explain why the window for full recall is so small. As with all 
other cases, the time window between full recall at $d=0$ and a single mis-identification $d=1$ was much 
greater than any other window between $d=n$ and $d=n+1$, which is consistent with Newton's Law of Cooling.

\subsection{Outlier: Acrylic Nails}
There was a single Stage 1 subject that had long acrylic fingernails. Instead of typing with fingertips, this person
tapped the keys with nail-tips. Since these do not have nearly as much surface area as fingertips, 
and false nails do not have any blood vessels to regulate their temperature, this subject left almost no thermal residue. 
In fact, not a single key-press could be correctly identified in any of the $40$ password entry trials. Consequently, 
this subject is not included in either  Touch or Hunt-and-Peck typist populations. However, as a side curiosity, we
note that, although it may be a rare occurrence, any user with long acrylic fingernails is virtually immune to \attacka.

\section{Discussion}
\label{sec:dis}
%What does this new side-channel vulnerability window mean for us?
%How can this be exploited
%What are the safeguards?
%
In this section, we break down our observations from Section \ref{sec:res} between  
two password classes, and among two categories of typists. % and note the strengths and weaknesses of our approach.

\subsection{Results with Common Passwords}
%English words are a lot easier to infer ordering
%numbers are still difficult
%creating a more accurate "distance" metric might help
Stage 2 subjects were particularly adept at identifying passwords that are English words or phrases. Even though we 
could not reliably detect the exact sequence of pressed keys, ordering can be found indirectly by mapping the set of 
%%GTS: what does it mean "we can provide" seems wishy-washy... Change!!!
%Okay -- introduce dictionary concept here, bring home later on
pressed keys to words (essentially, solving an anagram puzzle). Furthermore, a list of distances between 
detected keys (characters) and possible words, can be used to reconstruct full passwords from incomplete thermal residues.. 
Finally, the same list of distances can help determine when a key is pressed multiple times. These combinations 
highlight the threat posed by \attacka to already insecure passwords. 

\subsection{Results with Random Passwords}
%reduction in search space 
%we don't know when repeats occur
%we cannot find order
However, strong results from Stage 2 subjects' identification of English-language words does not extend to secure, 
randomly-selected passwords. First, inability to reliably determine the order of pressed keys can not be mitigated 
by leveraging the underlying linguistic structure. Moreover, it is unclear whether a given set of emanations represents the 
whole password, or if some information was lost. Finally, it is impossible to tell if a key was pressed multiple times. 
However, even with these shortcomings, our subjects managed to greatly reduce the password search space  
from $72^n$ to $72^{n-m}*m!$ where $n$ is the total number of characters in the password, and $m$ is the number 
of identified key-presses. 

\subsection{Results with Hunt-and-Peck Typists}
\begin{figure}[th!]
\fbox{\centering
\includegraphics[height=1.8in,width=\columnwidth]{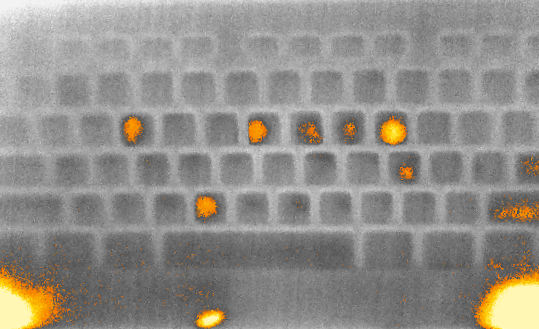} }
\caption{{\small Password "iloveyou" entered by a Hunt-and-Peck typist.}}
\label{fig:HandP}
\end{figure}
As described in Section \ref{subsec:handp}, Hunt-and-Peck typists are particularly vulnerable to \attacka. This is not surprising, 
given that these less-skilled typists tend to type more slowly, and primarily use their index fingers, which have greater fingertip 
surface area than ring or pinky fingers \cite{peters2009diminutive}. This results in greater heat transfer, due to longer contact 
duration with a larger contact area. Also, as seen from Figure \ref{fig:HandP}, Hunt-and-Peck typists do not touch
any keys that are not part of the password. Therefore, every observed key-press is part of the password.

\subsection{Results with Touch Typists}
\begin{figure}[th!]
\fbox{\centering
\includegraphics[height=1.8in,width=\columnwidth]{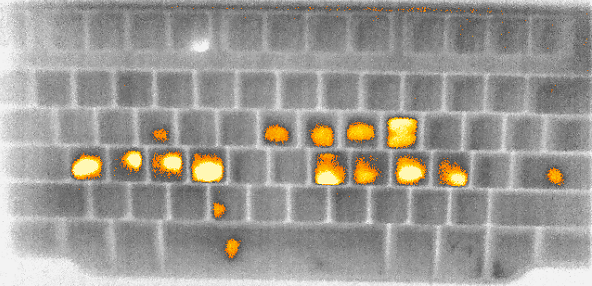} }
\caption{{\small Password "iloveyou" Entered by a Touch Typist.}}
\label{fig:home-row}
\end{figure}
For Touch typists, two factors confuse their thermal residues and make passwords harder to harvest. 
One is their habit to rest their hands on the home-row, which introduces potential false positives. as Figure \ref{fig:home-row} 
shows. This is exacerbated by the possibility that any home-row key might actually be part of the password. Because of this, 
stage 2 subjects were not penalized for classifying the home-row keys as pressed; they were instructed to identify all 
keys that looked to them as having been pressed. 

Another issue is that Touch typists tend to use all fingers of both hands while typing. This causes two advantages over
their Hunt-and-Peck counterparts. First, they touch individual keys for a shorter time, thus transferring less 
heat to the key-cap. Second, they type much more quickly and also use their ring and pinky fingers. 
Fingertips of these smaller fingers tend to have $1/2$ of the surface area of larger index or middle fingers. 
Thus, they transfer half of the total heat energy due to conduction during a key-press \cite{peters2009diminutive}. 
Such factors make Touch typists much more resistant to \attacka, particularly, 
at the level of our moderately sophisticated adversarial model.

\subsection{Ordering}
%
%TK: A lot of work needs to be done /w this subsection.
%TK: "In this section we" is unnessecary (too much "we")
%TK: Doesn't make sense to compare to RW (we have no context for it)
%l
\begin{figure*}[th!]
	\fbox{\centering
		\includegraphics[height=2.6in,width=1.75\columnwidth]{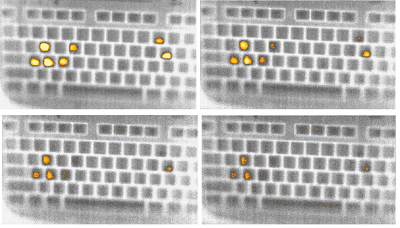} }
	\caption{{\small Password "passw0rd" thermal residue after 0(top left), 15 (top right), 
	30 (bottom left), and 45 (bottom right) seconds after entry}}
	\label{fig:timelapse}
\end{figure*}

Unfortunately, inspection of thermal images by stage 2 subjects did not yield any reliable key-press ordering information. 
Newton's Law of Cooling might seem to indicate that any reduction in heat energy would occur uniformly across 
all pressed keys, resulting in exposure of ordering. However, this is not true in practice. One reason is due to by 
\emph{keystroke inconsistency} in the dynamics of Touch typists. Factors, such as the travel distance between keys 
and the particular finger used to press a key, result in small differences in the duration, and total surface area of, contact. 
Since each key-press is distinct, intensity of a given thermal residue does not correspond to its relative position in the 
target password. This holds even for Hunt-and-Peck typists, who tend to use only their index fingers. As evidenced 
by Figure \ref{fig:timelapse}, Hunt-and-Peck typist does not necessarily press keys with uniform force or for a uniform duration. 
These inconsistencies make reliable ordering of key-presses infeasible in our analysis framework. 
However, as mentioned above, for insecure (language-based) passwords, dictionary tools can be used to infer the 
most likely key-press order.   

\subsection{Mitigation Strategies}
There are several simple strategies to mitigate or reduce the threat of \attacka, without modifying any existing hardware. 
The most intuitive solution is to introduce \emph{Chaff typing} right after a password is entered. This can be as simple 
as asking the users to swipe their hands along the keyboard after password entry, or requiring them to introduce noise 
by typing arbitrary ``chaff''. This would serve to obscure the password by introducing useless thermal residues, and thus
make the password key-presses much more difficult to retrieve. 
Another way is to avoid keyboard entry altogether and use the mouse to select (click on) password characters displayed
on the on-screen keyboard. A variation is to have drop-down menu for each position of the password and the user selects 
each character individually. A more burdensome alternative is to use the keyboard arrow keys to adjust a random 
character  string (displayed on the screen) to the actual password. All such methods are well-known and are quite viable. 
However, they are more vulnerable to \surfa, due to the ease of watching a victim's larger, visible screen instead of their smaller, 
partially occluded keyboard.
Finally, a user who is willing to go to extreme lengths to avoid leaving thermal residues could wear insulating gloves 
or rubber thimblettes over their fingers during password entry. This would greatly reduce thermal residues, 
and make \attack ineffective, since thermal conductivity of the insulating material would be much less than that of human skin. 

If hardware changes are possible, other mitigation techniques might be appropriate. 
For example, a touch-screen would allow password entry without the use of a keyboard. However, this  would be more
(than keyboard entry) vulnerable to \surfa. Also, the use of touch-screens opens the door for attacks that exploit 
smudge patterns left behind by fingers \cite{aviv2010smudge}.
Alternatively, common plastic keyboards could be replaced with metallic ones. Metals have much higher thermal conductivity 
than plastics. Thus, any localized thermal residues very quickly dissipate throughout the keyboard. A similar strategy was 
adopted to protect ATMs from thermal attacks \cite{mowery2011heat}.

\section{Comparison with Similar Attacks}
\label{sec:compare}
We now compare \attack with several similar human factors-based 
insider attacks. We focus on several aspects: adversary's \emph{Goal}, any \emph{Required Equipment}, the 
\emph{Timeliness} requirements, whether a \emph{Careless Victim} is needed, and finally, if \emph{Prior Profiling} of the
victim is required. Summary of the comparison is shown in Table \ref{tab:comp}.

\begin{table*}[th!]
\caption{Feature Comparison of Common Human-Based Attack Types.}
\label{tab:comp}
\centering
\begin{tabular}{|c||c|c|c|c|c|c|}
\hline\cline{1-6}
Attack              &     Attack                &   Adversary &  Careless  & Equipment & Prior Profiling
\\
Type:               &     Goal:                 &    Timeliness       &   Victim?    & Needed:  & Required?
\\ \hline \cline{1-6}
\lunch              &     Hijack  &     $15$ min (default)                  &   YES        & None & NO
\\ & Log-in Session  & & & \\ \hline
\surf                 &     Password          &    Real-Time                 &   YES        & Pair of Eyes  & NO
\\ & & & & or Video Camera&
\\ \hline
\acoust            &     Password          &    Real-Time                &   NO          & Audio & YES
\\ & & & & Recorder&
\\ \hline
\vibrat              &     Password          &     Real-Time               &   NO          & Accelerometer & YES
\\ \hline
\attack             &     Password          &     up to $1$ min                  &   NO          & Thermal & NO
\\ & & & & Camera&
\\ \hline \cline{1-6}
\end{tabular}
\end{table*}

\subsection{\lunch}
\luncha are performed by the insider adversary who relies on a careless victim that neglects to terminate 
their secure log-in session \cite{shamir1999playing}. 
\begin{compactitem}
\item \emph{Objective}: to gain access to a single secure (authenticated) session.
\item \emph{Required Equipment}: none, the adversary only needs to physically access the computer once the victim leaves. 
\item \emph{Timeliness:} determined by the de-authentication technique(s) used by the victim. For example, the default inactivity
timeout for Windows machines is a generous $15$ minutes. 
\item \emph{Careless Victim}: required for this attack to work. At the minimum, the victim needs to leave their 
workstation unattended without logging out or locking the screen.
\item \emph{Profiling}: no prior victim profiling is needed. The adversary can be opportunistic; it gains access to an 
authenticated session with out any additional or prior knowledge required.
\end{compactitem}

\subsection{\surf}
\surfa are performed by the insider adversary who looks over the shoulder of a careless victim while the password is entered.
It can also be performed with the aid of a (perhaps hidden) camera pointed at the victim's keyboard, in which case adversarial
presence is not required.
\begin{compactitem}
\item \emph{Objective}: to learn the victim's password. 
\item \emph{Required Equipment}: none, though a video camera can be useful.
\item \emph{Timeliness}  real-time, as the adversary must watch victim password entry as it occurs.
\item \emph{Careless Victim}: required, since the adversary has to stand over the victim's terminal to
 watch them type in their password. Careless victim is not required in case of a pre-placed viceoa camera.
\item \emph{Profiling}: no prior victim profiling is needed and the adversary can be opportunistic: it learns the 
victim's password with no additional or prior knowledge.
\end{compactitem}

\subsection{\acoust}
\acousta are performed by the insider adversary who instruments the victim's environment with an audio 
recording device and exploits acoustic dynamics \cite{zhuang2009keyboard}
\begin{compactitem}
\item \emph{Objective}: learn the victim's password. 
\item \emph{Required Equipment}: an audio recording device, placed nearby.
\item \emph{Timeliness}: real-time, since the adversary must record the keyboard sounds instantaneously.
\item \emph{Careless Victim}: not required; the recording device can be hidden from view.
\item \emph{Profiling}: prior victim profiling is needed; the adversary must build an acoustic profile 
of the victim to accurately interpret keystroke sounds.
\end{compactitem}

\subsection{\vibrat}
\vibrata are performed by the insider adversary using an accelerometer to record vibrations created by a victim 
typing into a keyboard, in order to reconstruct what was typed \cite{marquardt2011sp}.
\begin{compactitem}
\item \emph{Objective}: learn the victim's password. 
\item \emph{Required Equipment}: an accelerometer, placed nearby (closer than in \acousta).
\item \emph{Timeliness}: real-time, since the adversary must record the victim's vibrations instantaneously.
\item \emph{Careless Victim}: not required; the recording device can be hidden from view. 
\item \emph{Profiling}: prior victim profiling is needed; the adversary must build a vibration profile  in order
to accurately interpret keystroke vibration patterns.
\end{compactitem}

\subsection{\attack}
\attacka are performed by an insider adversary who records thermal residues users after recent password entry.
\begin{compactitem}
\item \emph{Objective}: learn the victim's password. 
\item \emph{Required Equipment}: thermal camera.
\item \emph{Timeliness}: up to $1$ minute, the adversary must record the keyboard before thermal residues dissipate.
\item \emph{Careless Victim}: not required; recording/imaging takes place after the victim leaves.
\item \emph{Profiling}: prior victim profiling not needed. The adversary does not need any prior knowledge of the victim
to analyze thermal images (though it obviously helps, especially with insecure passwords).
\end{compactitem}

\section{Related Work}
\label{sec:rw}
%%%GTS: This paragraph below is poor... Rewrite and re-read!!!!
%
Real-time attacks that target passwords (and countermeasures to them) have been studied extensively in the literature. 
Many methods have been proposed to 
mitigate \surfa \cite{brudy2014anyone, kumar2007reducing, yamamoto2009shoulder}. \cite{asonov2004keyboard, 
zhuang2009keyboard, berger2006dictionary, zhu2014context} have shown that keyboard acoustic emanations leak information 
about pressed keys. \cite{halevi2012closer} investigated how typing style (Hunt-and-Peck vs. Touch) influences keyboard acoustic 
emanation attacks.  As was recently shown, such attacks can be even mounted remotely \cite{compagno2017don}. 

The earliest attempt to use a thermal camera was focused on recovering key-codes entered 
into a rubber keypad of an industrial-grade safe \cite{SafeCracking}. Although not much detail is provided, 
it is argued that the attack can successfully yield key-codes up to $5-10$ minutes after initial entry.

Androitis et al. \cite{andriotis2013pilot} discuss using a thermal camera to infer screen-lock patterns of smartphones. This
study reports that screen-lock patterns can be seen up to $3$ seconds after entry when using a {\bf cold} just-booted 
smartphone. After a few seconds, it was no longer possible to extract any information about screen-lock 
patterns. In a similar effort, \cite{abdelrahman2017stay} conducted more extensive experiments with $18$ users to 
assess efficacy of thermal imaging attacks against screen-lock patterns. it was shown that PINs are 
vulnerable to such an approach, while swipe-patterns are not.

Mowery et al. \cite{mowery2011heat} investigated the influence of material composition (metal vs. plastic) and camera distance 
(14 vs. 28 inches) on PIN recovery, using a US\$$17,950$ thermal camera, on commercial PoS-style PIN pads. 
Results showed that metallic PIN pads are not prone to password recovery since  thermal residue dissipates rapidly and 
metallic surfaces partially reflect thermal energy. For plastic PIN pads, given a thermal camera placed $14$ inches away, 
$80$\% of pressed keys were correctly identified immediately after entry.  Success rate dropped down to $60$\% 
and $40$\% after $30$ and $60$ seconds, respectively. Perfect code recovery at any time is rather low: $<10$\%. 

\cite{sidhustudy} discuss the effectiveness of a low-cost thermal camera ($\approx US\$330$, attachable to a 
smartphone) to recover 4-digit PINs entered into rubber keypads. Analysis shows that the camera's distance 
from the keypad is important: from $22.9$ and $48.3$ centimeters, $2.6$ and $0.28$ digits can be identified, 
respectively up to $20$ seconds after entry.

Finally, \cite{wodo2016thermal} discusses viability of thermal imaging attacks on various PIN-entry devices. 
Analysis showed that the attack is a credible threat. In addition, the study discusses how metallic surfaces can 
be conditioned to make thermal imaging attack successful. Surface conditioning methods include: hair spray, 
stretch foil or transparent nail polish. Nail polish was reported to be the best, though success rates were not provided.

\section{Conclusions \& Future Work}
\label{sec:conc}
As formerly niche sensing devices become less and less expensive, new side-channel attacks move from "Mission: Impossible" 
towards reality. This strongly motivates exploration of novel human-factors attacks, such as those based on \attack. 
Work described in this paper sheds some light on understanding the thermodynamic relationship between human fingers 
and external computer keyboards. In particular, it exposes the vulnerability of standard password-based systems to 
adversarial collection of thermal emanations.

Based on the results of our study, we believe that \attacka represent a new credible threat for password-based systems, 
and that human-induced  thermal side-channels deserve further study. This is especially true considering the constantly decreasing 
cost and increasing availability of high-quality thermal imagers. To this end, we anticipate the following future work directions: 
\begin{figure}[ht!]
\fbox{\centering
\includegraphics[height=1.0in,width=0.7\columnwidth]{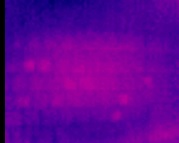} }
\caption{\small{"jordan23", $5$ seconds after entry, captured by TG165.}}
\label{fig:TG165}
\end{figure}
\begin{compactitem}
\item Given marked differences in collectible data between Touch and Hunt-and-Peck typists, one interesting next step is to 
further refine our attack approach to handle expert typists who introduce natural chaff through resting their hands on the 
keyboard home-row. Correct disambiguation of a home-row key being a part of the password rather than thermal noise,
would be very helpful in limiting the password search space. 
\item Another future direction is a longitudinal study to model multiple instances of \attacka, i.e., ,where the adversary, over time,
has several chances to obtain thermal imaging data against the same victim. Our study only measured thermal residues 
from each subject once, per password per keyboard. We hypothesize that a more persistent adversary would
be more successful and would be more likely to recover the entire password after multiple \attack instances. 
However, substantial further experiments are needed to substantiate this claim. 
\item It would also be useful to investigate lowering the bar for adversarial sophistication. 
Figure \ref{fig:TG165} shows an image of password {\sf ``jordan23''} entered by a Hunt-and-Peck typist, 
$5$ seconds after entry, as captured by the inexpensive FLIR TG165. 
Specifications for this camera are detailed in Section \ref{subsec:FLIRS}. This image 
suggests that, in the long run, even a less capable (in terms of equipment) adversary may pose a credible threat.
\item Finally, we intend to explore collected thermal data to find ordering effects on typed passwords. Currently, we can not 
correctly determine the sequence of pressed keys. However, this is probably a limitation of our specific equipment 
and not of the laws of thermodynamics. As shown in Section \ref{sec:bg} although the rate of cooling slows down markedly 
as a hot object approaches room temperature, there remains some heat difference that can be observed by a sensitive tool. 
Perhaps if we modify our approach to pick apart these differences, the overall strength of \attacka would be greatly increased. 
\end{compactitem}

\section*{Acknowledgements}
We would like to thank  Derek Dunn-Rankin and Michela Vicariotto for their generosity in lending us the FLIR SC620 for use in this study.
\bibliographystyle{acm}
\bibliography{hotstuff}

\begin{thebibliography}{10}

\bibitem{abdelrahman2017stay}
{\sc Abdelrahman, Y., Khamis, M., Schneegass, S., and Alt, F.}
\newblock Stay cool! understanding thermal attacks on mobile-based user
  authentication.
\newblock In {\em Proceedings of the 2017 CHI Conference on Human Factors in
  Computing Systems\/} (2017), ACM, pp.~3751--3763.

\bibitem{andriotis2013pilot}
{\sc Andriotis, P., Tryfonas, T., Oikonomou, G., and Yildiz, C.}
\newblock A pilot study on the security of pattern screen-lock methods and soft
  side channel attacks.
\newblock In {\em Proceedings of the sixth ACM conference on Security and
  privacy in wireless and mobile networks\/} (2013), ACM, pp.~1--6.

\bibitem{asonov2004keyboard}
{\sc Asonov, D., and Agrawal, R.}
\newblock Keyboard acoustic emanations.
\newblock In {\em Security and Privacy, 2004. Proceedings. 2004 IEEE Symposium
  on\/} (2004), IEEE, pp.~3--11.

\bibitem{aviv2010smudge}
{\sc Aviv, A.~J., Gibson, K.~L., Mossop, E., Blaze, M., and Smith, J.~M.}
\newblock Smudge attacks on smartphone touch screens.
\newblock {\em Woot 10\/} (2010), 1--7.

\bibitem{berger2006dictionary}
{\sc Berger, Y., Wool, A., and Yeredor, A.}
\newblock Dictionary attacks using keyboard acoustic emanations.
\newblock In {\em Proceedings of the 13th ACM conference on Computer and
  communications security\/} (2006), ACM, pp.~245--254.

\bibitem{brudy2014anyone}
{\sc Brudy, F., Ledo, D., Greenberg, S., and Butz, A.}
\newblock Is anyone looking? mitigating shoulder surfing on public displays
  through awareness and protection.
\newblock In {\em Proceedings of The International Symposium on Pervasive
  Displays\/} (2014), ACM, p.~1.

\bibitem{burton1939range}
{\sc Burton, A.}
\newblock The range and variability of the blood flow in the human fingers and
  the vasomotor regulation of body temperature.
\newblock {\em American Journal of Physiology-Legacy Content 127}, 3 (1939),
  437--453.

\bibitem{compagno2017don}
{\sc Compagno, A., Conti, M., Lain, D., and Tsudik, G.}
\newblock Don't skype \& type!: Acoustic eavesdropping in voice-over-ip.
\newblock In {\em Proceedings of the 2017 ACM on Asia Conference on Computer
  and Communications Security\/} (2017), ACM, pp.~703--715.

\bibitem{dai2004comparison}
{\sc Dai, T., Pikkula, B.~M., Wang, L.~V., and Anvari, B.}
\newblock Comparison of human skin opto-thermal response to near-infrared and
  visible laser irradiations: a theoretical investigation.
\newblock {\em Physics in Medicine \& Biology 49}, 21 (2004), 4861.

\bibitem{halevi2012closer}
{\sc Halevi, T., and Saxena, N.}
\newblock A closer look at keyboard acoustic emanations: random passwords,
  typing styles and decoding techniques.
\newblock In {\em Proceedings of the 7th ACM Symposium on Information, Computer
  and Communications Security\/} (2012), ACM, pp.~89--90.

\bibitem{kumar2007reducing}
{\sc Kumar, M., Garfinkel, T., Boneh, D., and Winograd, T.}
\newblock Reducing shoulder-surfing by using gaze-based password entry.
\newblock In {\em Proceedings of the 3rd symposium on Usable privacy and
  security\/} (2007), ACM, pp.~13--19.

\bibitem{marquardt2011sp}
{\sc Marquardt, P., Verma, A., Carter, H., and Traynor, P.}
\newblock (sp) iphone: decoding vibrations from nearby keyboards using mobile
  phone accelerometers.
\newblock In {\em Proceedings of the 18th ACM conference on Computer and
  communications security\/} (2011), ACM, pp.~551--562.

\bibitem{mickelberg2014us}
{\sc Mickelberg, K., Pollard, N., and Schive, L.}
\newblock Us cybercrime: rising risks, reduced readiness key findings from the
  2014 us state of cybercrime survey.
\newblock {\em US Secret Service, National Threat Assessment Center.
  Pricewaterhousecoopers\/} (2014).

\bibitem{mowery2011heat}
{\sc Mowery, K., Meiklejohn, S., and Savage, S.}
\newblock Heat of the moment: Characterizing the efficacy of thermal
  camera-based attacks.
\newblock In {\em Proceedings of the 5th USENIX conference on Offensive
  technologies\/} (2011), USENIX Association, pp.~6--6.

\bibitem{noyes1983qwerty}
{\sc Noyes, J.}
\newblock The qwerty keyboard: A review.
\newblock {\em International Journal of Man-Machine Studies 18}, 3 (1983),
  265--281.

\bibitem{occupational1999osha}
{\sc {Occupational Safety and Health Administration and others}}.
\newblock Osha technical manual.
\newblock {\em Section VIII\/} (1999).

\bibitem{owusu2012accessory}
{\sc Owusu, E., Han, J., Das, S., Perrig, A., and Zhang, J.}
\newblock Accessory: password inference using accelerometers on smartphones.
\newblock In {\em Proceedings of the Twelfth Workshop on Mobile Computing
  Systems \& Applications\/} (2012), ACM, p.~9.

\bibitem{peters2009diminutive}
{\sc Peters, R.~M., Hackeman, E., and Goldreich, D.}
\newblock Diminutive digits discern delicate details: fingertip size and the
  sex difference in tactile spatial acuity.
\newblock {\em Journal of Neuroscience 29}, 50 (2009), 15756--15761.

\bibitem{pyda2004heat}
{\sc Pyda, M., Nowak-Pyda, E., Mays, J., and Wunderlich, B.}
\newblock Heat capacity of poly (butylene terephthalate).
\newblock {\em Journal of Polymer Science Part B: Polymer Physics 42}, 23
  (2004), 4401--4411.

\bibitem{robb2014sony}
{\sc Robb, D.}
\newblock Sony hack: A timeline.
\newblock
  \url{http://deadline.com/2014/12/sony-hack-timeline-any-pascal-the-interview-north-korea-1201325501/},
  2014.

\bibitem{sauro2009estimating}
{\sc Sauro, J.}
\newblock Estimating productivity: composite operators for keystroke level
  modeling.
\newblock In {\em International Conference on Human-Computer Interaction\/}
  (2009), Springer, pp.~352--361.

\bibitem{shamir1999playing}
{\sc Shamir, A., and Van~Someren, N.}
\newblock Playing hide and seek with stored keys.
\newblock In {\em International conference on financial cryptography\/} (1999),
  Springer, pp.~118--124.

\bibitem{sidhustudy}
{\sc Sidhu, J.~S., Butakov, S., and Zavarsky, P.}
\newblock Study of potential attacks on rubber pin pads based on mobile thermal
  imaging.

\bibitem{wodo2016thermal}
{\sc Wodo, W., and Hanzlik, L.}
\newblock Thermal imaging attacks on keypad security systems.
\newblock In {\em SECRYPT\/} (2016), pp.~458--464.

\bibitem{yamamoto2009shoulder}
{\sc Yamamoto, T., Kojima, Y., and Nishigaki, M.}
\newblock A shoulder-surfing-resistant image-based authentication system with
  temporal indirect image selection.
\newblock In {\em Security and Management\/} (2009), pp.~188--194.

\bibitem{SafeCracking}
{\sc Zalewski, M.}
\newblock Cracking safes with thermal imaging.
\newblock \url{http://lcamtuf.coredump.cx/tsafe/}, 2005.
\newblock Accessed: 2018-04-02.

\bibitem{zhu2014context}
{\sc Zhu, T., Ma, Q., Zhang, S., and Liu, Y.}
\newblock Context-free attacks using keyboard acoustic emanations.
\newblock In {\em Proceedings of the 2014 ACM SIGSAC Conference on Computer and
  Communications Security\/} (2014), ACM, pp.~453--464.

\bibitem{zhuang2009keyboard}
{\sc Zhuang, L., Zhou, F., and Tygar, J.~D.}
\newblock Keyboard acoustic emanations revisited.
\newblock {\em ACM Transactions on Information and System Security (TISSEC)
  13}, 1 (2009), 3.

\end{thebibliography}
\end{document}